\documentclass[pra,aps,twocolumn,letterpaper]{revtex4-2}

\usepackage{color}
\usepackage{amsmath}
\usepackage{amssymb}  
\usepackage{bbold}
\usepackage{float}
\usepackage{pifont}   
\usepackage{graphicx} 
\usepackage{dcolumn}  
\usepackage{bm}       
\usepackage{amsfonts} 
\usepackage{multirow} 
\usepackage{breakurl}
\usepackage[hyperindex,breaklinks]{hyperref}
\usepackage{placeins}
\newcommand{\vect}[1]{\mathbf{#1}}
\newcommand{\ket}[1]{\left|{#1}\right\rangle}

 	\definecolor{lightblue}{rgb}{0.120,0.544,0.916}
 	\definecolor{redd}{rgb}{0.864, 0.108, 0.384}
\begin{document}

\renewcommand\floatpagefraction{0.88} 
\renewcommand\topfraction{0.88}       

\author{Michael Vogl$^{1,2}$}
\author{Pontus Laurell$^3$}
\author{Hao Zhang$^{4,5}$}
\author{Satoshi Okamoto$^5$}
\author{Gregory A. Fiete$^{6,7}$}
\affiliation{$^1$Department of Physics, King Fahd University of Petroleum and Minerals, 31261 Dhahran, Saudi Arabia}
\affiliation{$^2$Department of Physics, The University of Texas at Austin, Austin, TX 78712, USA}
\affiliation{$^3$Center for Nanophase Materials Sciences, Oak Ridge National Laboratory, Oak Ridge, Tennessee 37831, USA}
\affiliation{$^4$Department of Physics and Astronomy, University of Tennessee, Knoxville, TN 37996-1200, USA}
\affiliation{$^5$Materials Science and Technology Division, Oak Ridge National Laboratory,
	Oak Ridge, TN 37831, USA}
\affiliation{$^6$Department of Physics, Northeastern University, Boston, MA 02115, USA}
\affiliation{$^7$Department of Physics, Massachusetts Institute of Technology, Cambridge, MA 02139, USA}

\title{Exact resummation of the Holstein-Primakoff expansion and differential equation approach to operator square-roots}
\date{\today}

\begin{abstract}
Operator square-roots are ubiquitous in theoretical physics. They appear, for example, in the Holstein-Primakoff representation of spin operators and in the Klein-Gordon equation. Often the use of a perturbative expansion is the only recourse when dealing with them. In this work we show that under certain conditions differential equations can be derived which can be used to find perturbatively inaccessible approximations to operator square-roots. Specifically, for the number operator $\hat n=\hat a^\dag a$ we show that the square-root $\sqrt{\hat n}$ near $\hat n=0$ can be approximated by a polynomial in $\hat n$. This result is unexpected because a Taylor expansion fails. A polynomial expression in $\hat n$ is possible because $\hat n$ is an operator, and its constituents $a$ and $a^\dag$ have a non-trivial commutator $[a,a^\dag]=1$ and do not behave as scalars. We apply our approach to the zero mass Klein-Gordon Hamiltonian in a constant magnetic field, and as a main application, the Holstein-Primakoff representation of spin operators, where we are able to find new expressions that are polynomial in bosonic operators. We prove that these new expressions exactly reproduce spin operators. Our expressions are manifestly Hermitian, which offer an advantage over other methods, such as the Dyson-Maleev representation.
\end{abstract}


\maketitle

\section{Notice of copyright}
This manuscript has been authored in part by UT-Battelle, LLC, under contract DE-AC05-00OR22725 with the US Department of Energy (DOE). The US government retains and the publisher, by accepting the article for publication, acknowledges that the US government retains a nonexclusive, paid-up, irrevocable, worldwide license to publish or reproduce the published form of this manuscript, or allow others to do so, for US government purposes. DOE will provide public access to these results of federally sponsored research in accordance with the DOE Public Access Plan (\url{http://energy.gov/downloads/doe-public-access-plan}).


\section{Introduction}
Square-roots of operators appear in a large number of contexts in theoretical physics, and also play an important role in operator theory. In some cases, it is practical to calculate the operator square-root (OSR) using explicit formulas or by diagonalizing the operator. Often, however, there is only a very limited set of analytical tools to treat them, typically in the form of perturbative expansions.

This is not because OSRs represent a niche problem. Indeed one of the earliest appearances was near the beginning of quantum mechanics in the square-root of the Klein-Gordon equation \cite{1926ZPhy...37..895K,1926ZPhy...40..117G,doi:10.1063/1.1703882,doi:10.1063/1.530015,namsrai1998square,shakeri2008numerical,HAAS_2013}. Even for such an old problem it may prove useful to have a larger analytical toolbox. Another prominent example of OSRs occurs in the Holstein-Primakoff spin representation \cite{PhysRev.58.1098,Auerbach1994}, which is the usual starting point for spin-wave theory calculations. A third important OSR shows up in the context of quantum information in the form of the fidelity function \cite{Nielsen2010,GU_2010,bengtsson2017geometry,jozsa1994fidelity,barnum1996noncommuting,Raginsky_2001,Peters_2004,_yczkowski_2005,zanardi2007mixed,Paunkovi__2008,Mendon_a_2008,Wang_2009,Quan_2009,Marian_2012} and the Bures metric \cite{bures1969extension,uhlmann1976transition,_yczkowski_2005,Mendon_a_2008,Marian_2012,bengtsson2017geometry}, both used to quantify the closeness of two quantum states. The purpose of the current paper is, however, not to review all examples of OSRs, but to introduce a non-perturbative approximation of OSRs.

Our method is inspired by several flow equation approaches to many-body problems. For instance, the Wegner flow equation approach \cite{Wegner1994}, which was applied to various problems \cite{Wegner1994,Kehrein2007,PhysRevB.97.060201,quito2016localization,bach2010rigorous,wegner2001flow,mielke1998flow,lenz1996flow,kehrein1995flow,wegner2006flow,gubankova1998flow,ragwitz1999flow,kehrein1994flow,wegner1998flow,Kelly_2020}, allows for a non-perturbative diagonalization of a Hamiltonian using flow equations for its couplings. In this approach, the problem of diagonalization is recast in terms of differential equations. Similar methods have recently been used by some of us to find effective Floquet Hamiltonians \cite{Vogl_2019}, and various approximations to the time evolution operator \cite{Vogl_2019HJ}. Differential equation approaches have also been used in the method of unitary integration for the Liouville-Bloch equation \cite{PhysRevLett.81.4785} and Lindblad equation \cite{Rau_2002}.  We aim to use a similar approach to approximate an operator square-root.

The application that may be of most current interest is the Holstein-Primakoff (HP) OSR. The HP representation is typically used in the context of spin (local moment) models to represent deviations around a well-defined spin order in terms of a single species of boson per lattice site. It allows for a perturbative expansion in the number operator of such bosons, and ultimately leads to linear \cite{toth2015linear} and non-linear \cite{RevModPhys.85.219} spinwave descriptions of quantum magnets \cite{Auerbach1994}. However, for many systems of interest a ground-state spin ordering may be unknown or fail to exist, such as in frustrated systems \cite{Schmidt2017}, spin liquids \cite{balents2010spin,Broholm_2020,Zhou_2017,Knolle_2019}, and one-dimensional systems \cite{Giamarchi2004}. In such cases, the perturbative expansion often proves inaccurate or inconvenient.

Instead, more symmetric spin representations such as Schwinger bosons \cite{Auerbach1994,auerbach2011schwinger} and slave particle approaches \cite{RevModPhys.78.17,Shindou_2009} are commonly used, but require the introduction of auxiliary fields. Other fermionic approaches include the Jordan-Wigner representation of spin-$1/2$ operators \cite{1928ZPhy...47..631J}, and its generally complicated-to-use generalizations to higher dimensions \cite{PhysRevLett.63.322,PhysRevLett.71.3622} and higher spin \cite{PhysRevLett.86.1082,Dobrov_2003,PhysRevB.71.092404}. An important, equivalent alternative to the HP representation that also uses a single boson species, but avoids the square-root, is the Dyson-Maleev representation \cite{PhysRev.102.1217,Maleev1958,Itoi_1994,RevModPhys.63.375}. However, it has the drawback of generically breaking hermiticity. This is by no means an exhaustive list of spin representations --- indeed, other representations can be be found in Refs.\cite{Villain1974,Villain1975,PhysRevB.19.4780,10.1088/0305-4470/13/2/014,GARBACZEWSKI197865,Zhou_1999}. Since each of the available approaches has its own unique advantages and drawbacks, we will in this paper derive expressions for spin operators that i) involve only one boson species satisfying the canonical bosonic commutation relation, ii) preserve hermiticity, and iii) do not include square-roots of operators or other non-polynomial functions of operators.

Some of the expressions we derive were previously found to finite order in Ref.~\cite{Lindgard1974,batyev1986antiferromagnet} by a matching matrix elements (MME) method and have also been usefully applied in \cite{PhysRevB.93.224402} to capture effects beyond the reach of a $1/S$ expansion. Unlike the normal Taylor expansion of the HP OSR, the MME expansion and our result are able to correctly describe the symmetry in a Heisenberg model with easy-plane anisotropy as we will see later in the text. This is a long standing problem and was discussed using a slightly different approach in \cite{Tsuru1986}.  Our expansion thus naturally captures the same physics. However, unlike previous works, we present results to all orders and show that the expressions are exact when truncated to an appropriate order that depends on the spin length $S$. This feature was missed in all previous discussions we are aware of, since they focused entirely on reproducing commutation relations of spin operators. We, however, use a slightly softer exactness criterion. Namely, we require only that the operators are block-diagonal with physical and unphysical subspace blocks. In the commutator language we require that the commutators are reproduced up to a term that acts exclusively on the unphysical subspace, without coupling to the physical subspace. This is akin to allowing an inaccessible ``dark sector'' in the spin operator algebra. A more detailed discussion of this rationale is given in the main text.

Our hope is that such a representation may prove useful in describing spectral features not readily captured by conventional spin-wave theory, as is the case in e.g. the triangular-lattice antiferromagnet Ba$_3$CoSb$_2$O$_9$ \cite{Kamiya2018,PhysRevB.93.224402}, and quantum spin liquid candidates. Among the latter, the Kitaev spin liquid \cite{Kitaev_2006,Yang_2008,Vidal_2008,Schmitt_2015,willans2011site,pedrocchi2011physical,burnell20112,janvsa2018observation,gorshkov2013kitaev,halasz2016resonant,hickey2019emergence,wang2010reduced,cui2010quantum,halasz2014doping,wang2010realization,abasto2009thermal,schmoll2017kitaev,bolukbasi2012rigorous,kells2009finite,dusuel2008perturbative} is receiving particularly intense attention, since it hosts anyonic excitations of interest to topological quantum computing. While the ideal model is solvable \cite{Kitaev_2006}, and its dynamics known \cite{PhysRevLett.112.207203}, the description of realistic candidate materials \cite{Takagi2019,doi:10.7566/JPSJ.89.012002,trebst2017kitaev} require additional Hamiltonian terms, which generically breaks integrability. Some such candidates include $\alpha$-RuCl$_3$ \cite{kim2015kitaev,Banerjee2016,sandilands2016spin,Banerjee2017,Do2017,ran2017spin,glamazda2017relation,wolter2017field,Banerjee2018,yu2018ultralow,lampen2018anisotropic,Balz2019,eichstaedt2019deriving,laurell2020dynamical},
 CrI$_3$ \cite{xu2018interplay,stavropoulos2019microscopic,lee2020fundamental,aguilera2020topological,rodriguez2020phonon} and honeycomb iridium oxides \cite{chaloupka2010kitaev,kimchi2011kitaev,singh2012relevance,simutis2018chemical}.

The manuscript is structured as follows. In the next section of the paper we discuss how to compute square-roots of operators by using a differential equation approach. In section \ref{expansion_sqrtn} we show how this formalism may be used to find a series expansion for $\sqrt{a^\dag a}$ near $a^\dag a\approx 0$ in terms of integer powers of $(a^\dag a)$, which is an unexpected result because $\sqrt{x}$ cannot be expanded in integer powers of $x$ near $x=0$. Of course, since $a^\dag a$ is an operator $a^\dag a\approx 0$ is a shorthand for "in the part of the Hilbert space where where matrix elements are close to zero". We will use similar shorthands throughout the text. This shows that a Taylor series may not always be ideal for finding power series expansions of operator functions. In section \ref{sec:KG} we then apply the method to the Klein-Gordon particle in a magnetic field with small or zero-mass --- such as in graphene. In section \ref{sec:HP} we present our main application to the Holstein-Primakoff representation of spin operators. We stress that the results we obtain are exact expressions for spin operators that are polynomial in bosonic operators. Lastly we present our conclusion.

\section{General Formalism}
The goal of this section is to find an operator differential equation that can be used to calculate a square root of two operators, $\sqrt{O_1+O_2}$, where $O_1$ and $O_2$ are both operators defined on the same complex Hilbert space $\mathcal{H}$.  We will make two simplifying assumptions. First, we assume that a square-root of one of the operators, $O_1$, is known or easy to calculate. Second, we assume that the two operators commute, $[O_1,O_2]=0$. Both these assumptions also have to be made in order for a Taylor expansion in $O_2$ to be viable (the more generic case is more involved, see Appendix~\ref{general_expansion} for details). For instance one could have $O_1=c\mathbb{1}$ with $c\in \mathbb{C}$, and $O_2$ any other operator. It should be noted that the OSR of an operator $O_1$ can have multiple branches, which may seem like an ambiguity. However, the choice of branch will be encoded in the initial conditions for the differential equations we derive, and should be informed by the problem at hand. Different branch choices can lead to different physics --- e.g. a branch with complex eigenvalues could not be used to describe a Hermitian Hamiltonian. The way we will compute $\sqrt{O_1+O_2}$ is by introducing the second operator $O_2$ in infinitesimal steps. To keep track of the steps we introduce a dummy parameter $s$ and define
\begin{equation}
	O_{\sqrt{\;}}(s):=\sqrt{O_1+sO_2}.
\end{equation}

Using the assumption $[O_1,O_2]=0$ we find that sending $s\to s+\delta s$ gives
\begin{equation}
	O_{\sqrt{\;}}(s+\delta s)=O_{\sqrt{\;}}(s)+\frac{\delta s}{2O_{\sqrt{\;}}(s)}O_2
\end{equation}
if $\delta s$ is infinitesimal and we therefore did a Taylor expansion of the right hand side.

A Taylor expansion of the left side gives us the differential equation
\begin{equation}
	\frac{d O_{\sqrt{\;}}(s)}{ds}=\frac{1}{2O_{\sqrt{\;}}(s)}O_2
	\label{cumbersomeDGL}
\end{equation}
that makes it possible to find $O_{\sqrt{\;}}(s)$ by introducing $O_2$ via infinitesimal steps. Note that this also means that $\sqrt{O_1}$ for the branch used needs to be invertible, or at least the limit of an invertible operator as we will see in the upcoming section.

The issue with this equation is that calculating the inverse of an operator is difficult. That is, we cannot easily make an ansatz for $O_{\sqrt{\;}}(s)=\sum_n c_n(s)\hat O_n$ as a sum of operators with $s$-dependent coefficients and solve this equation because calculating the inverse of the ansatz is difficult. 

This issue can be resolved with a little bit of extra work. We define
\begin{equation}
	O_{\sqrt{\;}}^{-1}(s):=\frac{1}{O_{\sqrt{\;}}(s)}.
	\label{cumbersome_variation}
\end{equation}
In this case Eq.~\eqref{cumbersomeDGL} becomes
\begin{equation}
	\frac{d O_{\sqrt{\;}}(s)}{ds}=\frac{1}{2}O_{\sqrt{\;}}^{-1}(s)O_2,
	\label{rel_sqrttoderiv}
\end{equation}
and we now need to find a differential equation for $O_{\sqrt{\;}}^{{-1}}(s)$, which can be obtained by Taylor expanding $O_{\sqrt{\;}}^{{-1}}(s+\delta s)$ in a similar way to above,
\begin{equation}
	\frac{dO_{\sqrt{\;}}^{-1}( s)}{ds}=-\frac{1}{2}(O_{\sqrt{\;}}^{-1}(s))^3 O_2.
	\label{diffforinversesqrt}
\end{equation}
One may insert Eq.~\eqref{rel_sqrttoderiv} in Eq.~\eqref{diffforinversesqrt}, and we find after rearranging that
\begin{equation}
	\frac{1}{2}O_2 \frac{d^2O_{\sqrt{\;}}(s)}{ds^2}=-\left(\frac{dO_{\sqrt{\;}}(s)}{ds}\right)^3.
	\label{good_dgl}
\end{equation}
The equation in this form is now useful to find coefficients $C_i$ for an ansatz $O_{\sqrt{\;}}(s)=\sum_n  c_n(s)\hat O_n$ because powers of this operator are trivial to compute.

\section{Expanding the square-root of the number operator}
\label{expansion_sqrtn}

We may now use equation \eqref{good_dgl} to find an expansion of 
$\sqrt{a^\dag a}$. In the language of the previous section for this case $O_2=a^\dag a$ and $O_1=0^+\mathbb{1}$, where $0^+$ signifies a dummy variable that eventually will take a directed limit to zero. One can make the ansatz 
\begin{equation}
	\sqrt{sa^\dag a}\approx \sum_n C_n(s)(a^\dag)^n a^n
	\label{ansatz}
\end{equation}  and compare coefficients of $(a^\dag)^n a^n$ to find a set of differential equations for $C_n$.  If we truncate at third order we find
\begin{widetext}
	\begin{equation}
	\begin{aligned}
	C_0^\prime&=0\\
	\frac{C_1''(s)}{2}&=-C_1'(s)^3\\
	\frac{C_2''(s)}{2}&=-6 C_1'(s) C_2'(s) \left[C_1'(s)+C_2'(s)\right]-C_1'(s)^3-2 C_2'(s)^3\\
	\frac{C_3''(s)}{4}&=-36 C_2'(s) C_3'(s) \left[C_1'(s)+C_2'(s)+C_3'(s)\right]-3 C_1'(s) C_2'(s) \left[C_1'(s)+4 C_2'(s)\right]\\
	&-9 C_1'(s) C_3'(s) \left[C_1'(s)+2 C_3'(s)\right]-10 C_2'(s)^3-12 C_3'(s)^3
	\end{aligned}
	\label{diffeqeqsqrt}
	\end{equation}
\end{widetext}
and initial conditions 
\begin{equation}
	C_{0,1,2,3}(0)=0;\quad C_1'(0)=\frac{1}{2\sqrt{0^+}};\quad C_{2,3}^\prime(0)=0.
	\label{initSQRT}
\end{equation}
The initial conditions were found by comparison to the infinitesimal case that is accurately described by a first order Taylor series. Note that the term with $\frac{1}{2\sqrt{0^+}}$ represents a directional limit that has to be taken at the end but $0^+$ can first be replaced by a dummy variable.

If we solve the equations and set $s=1$ and take the limit for $0^+$ we find that
\begin{equation}
	\sqrt{a^\dag a}\approx a^\dag a+\frac{\sqrt{2}-2}{2}a^{\dag ^2} a^2+\frac{3-3 \sqrt{2}+\sqrt{3}}{6} a^{\dag ^3} a^3.
\end{equation}

One should note  that this expression can be put in terms of powers of $\hat n= a^\dag  a$ and is valid near $ a^\dag  a=0$. 
More precisely, in what sense does this expansion converge to the correct operator? The answer is that by including terms up to $\left( a^{\dag}\right)^n a^n$ the $n+1$ lowest eigenvalues are exactly reproduced; higher eigenvalues are approximated more accurately as well.

It is important to stress that the square-root $\sqrt{x}$ is non-analytic near $x=0$. Yet we were able to find an expansion in terms of powers of $x$ that is valid near $x=0$. 

\section{Application to the Klein-Gordon square-root}\label{sec:KG}
The method for finding a non-perturbative expansion of an operator square-root can of course also be used for the Klein-Gordon square-root Hamiltonian for relativistic particles. Let us for instance consider the 2D Hamiltonian
\begin{equation}
 H=\sqrt{m^2+p^2}+V(x,y).
\end{equation}

If this system is subjected to a constant magnetic field given by $\vect A=\frac{B}{2}(-y,x)$ one may introduce the magnetic field by minimal substitution $p_i\to\Pi_i=p_i-A_i$ and one can introduce creation and annihilation operators, $a=\sqrt{\frac{1}{2B}}(\Pi_x+i\Pi_y)$, to find the Hamiltonian
\begin{equation}
 H=\sqrt{4|B|S}\sqrt{1+\frac{a^\dag a}{2S}}+V(x,y),
\end{equation}
where we introduced a short-hand $S=\frac{m^2+|B|}{4|B|}$. The operator now bears a striking resemblance to the square-root that appears in the Holstein-Primakoff spin representation, which we will discuss later. A straightforward Taylor expansion of the square-root in terms of $1/S$ already yields corrections
\begin{equation}
 H\approx \sqrt{4|B|S}-\frac{1}{4}\sqrt{\frac{|B|}{S}}+\frac{1}{4}\sqrt{\frac{1}{|B|S}}\Pi^2+V(x,y)
\end{equation}
to what one would expect from the non-relativistic limit of large mass
\begin{equation}
\sqrt{m^2+\Pi^2}\approx m+\frac{\Pi^2}{2m}+V(x,y).
\end{equation}
This approximation lifts the restrictions to large masses from the non-relativistic limit as long as one considers strong magnetic fields. 

However, we can do better without the introduction of further complications. That is, we can make the ansatz $\sqrt{1+sa^\dag a}=\sum_n C_n(s)(a^\dag)^n a^n$, which means that we can employ the first two differential equations from \eqref{diffeqeqsqrt}  to approximate the square root. For this we have to choose slightly different initial conditions than previously, $C_0(0)=1$, $C_1(0)=0$, $C_1'(0)=1/2$ and let s run up to $s=1/(2S)$. The initial conditions are again found by comparison to a first order Taylor expansion. The result we find is
\begin{equation}
\begin{aligned}
& H\approx \sqrt{4|B|S}\left[1+\left(\sqrt{1+\frac{1}{2S}}-1\right)\frac{S}{2|B|}\Pi^2\right]+V(x,y)
\end{aligned}.
\hspace*{-0.5cm}
\label{approx_kgsqrt}
\end{equation}
This new approximation is now more reliable for small $|B|$, $m$ and level number $n$. This is seen most easily in the case of $V(x,y)=0$ where it is easy to check that it reproduces the lowest two energy levels $n=0,1$ exactly (recall that $\hat n=a^\dag a=S\Pi^2/(2|B|)$).

The advantage of this approximation over an exact solution is that a quadratic $V(x,y)$ can be added and an analytic solution of this approximate problem is still possible because this is still a harmonic oscillator. Note that in this case we would be able to find an approximation that is non-perturbative in $1/S$.

\section{Resummed Holstein-Primakoff expansion}\label{sec:HP}
We will now turn to our most interesting application --- an expansion for the square-root in the Holstein-Primakoff representation of a spin operator.
\subsection{Review of the method}
The Holstein-Primakoff representation \cite{PhysRev.58.1098} of spin-$S$ operators is given as
\begin{equation}
\begin{aligned}
&S^+=\hbar \sqrt{2S} \sqrt{1-\frac{a^\dagger a}{2S}}\, a\\
	&S^- = \hbar \sqrt{2S} a^\dagger\, \sqrt{1-\frac{a^\dagger a}{2S}}\\
		&S^z = \hbar(S - a^\dagger a)
\end{aligned}.
\label{holstein-primakoff}
\end{equation}
A few notes are due. For finite $S$ only finitely many bosonic excitations correspond to physical states. That is, bosonic excitations correspond to spin projections i.e. $S^z$ can only take eigenvalues in $\{ -S, -S+1, \dots ,S\}$. Hence, for spin $S$ we have the restriction $a^\dagger a \leq 2S$, which is also signaled by the fact that the square-root becomes imaginary for higher occupation numbers. 
This means that the Hilbert space is a Fock space, $F(\mathcal{H})=\bigoplus_{n=0}^\infty \mathcal{S}\mathcal{H}^{\otimes n}$, where $\mathcal{S}$ is the symmetrization operator, $\mathcal{H}^{\otimes n}$ denotes $n$ tensor products of the single particle Hilbert space $\mathcal{H}$. For spin $S$ the physical part of the Hilbert space is restricted such that it has the basis $\{\ket{0},...,\ket{2S}\}$.

\subsection{Exactness of the Holstein Primakoff approximation}
To see that the Holstein-Primakoff representation is an exact description of spin operators it is enough to check that it fulfills the correct spin algebra. 
For instance $[S^+,S^-]=2S^z$.

This reasoning is slightly restrictive so let us soften it a bit. The key feature of the Holstein Primakoff representation is that the spin operators $S^{+,-,z}$ reproduce the exact spin operators on the physical part of the Hilbert space and at the same time have no elements that couple to the unphysical part of the Hilbert space.
That is, in the occupation basis they have the form
\begin{equation}
	S^{+,-,z}=\begin{pmatrix}
		S^{+,-,z}_{phys}&0\\
	0&	S^{+,-,z}_{unphys}
	\end{pmatrix}.
	\label{Holstein_spin_op_block}
\end{equation}
In particular, for $S=1/2$ the explicit form of $S^+$ in the occupation basis is
\begin{align}
	S^+	&=	\left( \begin{array}{cc|ccccc}
		\color{lightblue}0	& \color{lightblue} 1	& & & & &\\
		\color{lightblue}0	& \color{lightblue} 0	& & & & &\\\hline\rule{0pt}{2.6ex}
		&& \color{redd}0 &\color{redd}i\sqrt{3}&\color{redd}0&\color{redd}\cdots&\color{redd}0\\
		&& \color{redd}\vdots&&\color{redd}\ddots&&\color{redd}\vdots\\
		&& \color{redd}0&\color{redd}\cdots&&&\color{redd}0
			\end{array}\right).
\end{align}

One sees that it splits into the physical (highlighted in blue) and unphysical (red) Hilbert spaces like in equation \eqref{Holstein_spin_op_block}.
The physical block is just the conventional $S^+$ matrix for spin 1/2. Importantly there is no coupling between physical and unphysical parts of the Hilbert space. This is what makes the method exact. 

Note that, because of this block structure, a spin Hamiltonian exactly written in the bosonic language will also separate into physical and unphysical blocks because the product of block diagonal matrices stays block diagonal. That is, the Hamiltonian is block diagonal of the form
\begin{equation}
	H=\begin{pmatrix}
	H_{phys}&0\\
	0&H_{unphys}
	\end{pmatrix}.
\end{equation}
One now can see that diagonalising the Hamiltonian one will find the exact physical eigenvalues and spurious unphysical ones.
\subsection{Usual Approach: Taylor expansion}
While the expressions in Eq. \eqref{holstein-primakoff} provide an exact way to represent the spin operators this is not too useful by itself because the square-roots are impractical to work with. One usually does a Taylor expansion around large $S$, using $1/S$ as expansion parameter.
\begin{equation}
	\begin{aligned}
	S^+\approx \hbar \sqrt{2S}\left(1-\frac{1}{4S}a^\dag a-\frac{1}{32S^2}(a^\dag a+a^{\dag^2}a^2)\right.\\
	-\left.\frac{1}{128 S^3}(a^{\dagger } a+3 a^{\dagger ^2} a^2+a^{\dagger^3 } a^3)\right)a.
	\end{aligned}
	\label{Taylor_sp}
\end{equation}
 This approach is most often also used in the case of $S=\frac{1}{2}$, where it is slightly surprising that it is justified. To see why it is justified recall that as mentioned above for smaller spins only states with few bosonic excitations e.g. $\{|0\rangle,|1\rangle\}$ for spin $\frac{1}{2}$ are physical. Therefore  acting in this part of the Hilbert space $a^\dag a|_{phys}\leq 1$ and the expansion is valid.
 
Although the expansion is useful it is not exact when truncated at any finite order. 
The spin operators $S^{+,-}$ no longer separate into physical and unphysical blocks, but couple physical and unphysical parts of the Hilbert space. For example, for spin 1/2 the spin operator $S^+$ in Eq.~\eqref{Taylor_sp} has the form

\begin{align}
S^+	&\approx	\left( \begin{array}{cc|ccccc}
\color{lightblue}0	& \color{lightblue} 1	& & & & &\\
\color{lightblue}0	& \color{lightblue} 0	& \frac{5}{8\sqrt{2}} & & & &\\\hline\rule{0pt}{2.6ex}
&& \color{redd}0 &\color{redd}i\sqrt{3}&\color{redd}0&\color{redd}\cdots&\color{redd}0\\
&& \color{redd}\vdots&&\color{redd}\ddots&&\color{redd}\vdots\\
&& \color{redd}0&\color{redd}\cdots&&&\color{redd}0
\end{array}\right).
\end{align}

One may see that physical and unphysical parts of the Hilbert space get coupled by the term $\frac{5}{8\sqrt{2}}$.

Generically a spin Hamiltonian using this approximate bosonic language when expressed in occupation number space has the form

\begin{equation}
	H=\begin{pmatrix}
	H_{phys}&\Delta\\
	\Delta^\dag &H_{unphys}
	\end{pmatrix},
\end{equation}

where $\Delta$ is the small coupling between physical and unphyical parts of the Hilbert space. It leads to unphysical contributions in the physical eigenvalues. The method is not exact anymore.

 \subsection{Improved Expansion}
As mentioned we can improve on the expansion. One may use the differential equation \eqref{good_dgl} to find such an improved expansion of the square-root. Like previously one may use the ansatz \eqref{ansatz} to introduce $-\frac{a^\dagger a}{2S}$ by infinitesimal steps. However, because we need to decrease terms under the square-root rather than increase them, one has to replace $d/ds\to -d/ds$ in \eqref{diffeqeqsqrt}. The second thing that changes compared to before are two of the initial conditions
 \begin{equation}
 	\begin{aligned}
 	C_0(0)=1;\quad C_1'(0)=-\frac{1}{4},
 	\end{aligned}
 \end{equation}
 while the other initial conditions in \eqref{initSQRT} remain unchanged.
 
 The solution to these differential equations \eqref{diffeqeqsqrt} for $s=\frac{1}{S}$ with the new initial conditions gives us an improved Holstein-Primakoff expansion up to third order, which we will not present here.
 
 Rather with additional work one may find that it is possible to construct higher order terms by the same scheme. After analysing additional orders one can see a pattern emerge. We find that the full expansion is given as
 \begin{equation}
 	\begin{aligned}
 		&S_+\approx \hbar \sqrt{2S}\left[\sum_{n=0}^{n_{\mathrm{max}}} Q_n a^{\dag^n} a^n\right]a;\quad Q_0=1,\\ &Q_n=\frac{1}{n!}A_n-\sum_{m=0}^{n-1}\frac{1}{(n-m)!}Q_m;\quad A_n=\sqrt{1-\frac{n}{2S}},
 	\end{aligned}
 	\label{new_expans_spin}
 \end{equation}
 where we prove later that this amounts to exact expressions for spin operators. It should be noted that during the review process of this manuscript equivalent expressions in closed form were also found by elegant alternative means via a Newton series expansion \cite{konig2020newton}.

Let us for now truncate at $n_{\mathrm{max}}=1$ to find
 \begin{equation}
\begin{aligned}
&S^+\approx \hbar \sqrt{2S}\left[1+\left(\sqrt{1-\frac{1}{2 S}}-1\right)a^\dag a\right]a,
\end{aligned}
\label{trunc_splus}
\end{equation}
 and discuss the case of spin $S=\frac{1}{2}$ to most easily see what kind of improvement we achieved. One may note that an expansion around large $S$ gives back the results for the Taylor expansion. In that sense our new expansion is a resummation of the Taylor series.

In the occupation basis  we find
\begin{align}
 	S^+	&=	\left( \begin{array}{cc|ccccc}
 	\color{lightblue}0	& \color{lightblue} 1	& & & & &\\
 	\color{lightblue}0	& \color{lightblue} 0	& & & & &\\\hline\rule{0pt}{2.6ex}
 	&& \color{redd}0 &\color{redd}-\sqrt{3}&\color{redd}0&\color{redd}\cdots&\color{redd}0\\
 	&& \color{redd}\vdots&&\color{redd}\ddots&&\color{redd}\vdots\\
 	&& \color{redd}0&\color{redd}\cdots&&&\color{redd}0
 	\end{array}\right).	\label{eq:splusnew:matrix}
\end{align}
Therefore the spin operator reproduces the physical matrix elements of $S^+$, and the physical block does not couple to the unphysical block like in Eq. \eqref{Holstein_spin_op_block}. 
One can show also more explicitly that there are no coupling between physical and unphysical parts of the Hilbert space
\begin{equation}
\begin{aligned}
&\langle 0| S^+|1\rangle=\hbar,\\
&\langle n\neq 0| S^+|1\rangle=\langle n| S^+|0\rangle=\langle 0|S^+|n\neq 1\rangle=0.
\end{aligned}
\end{equation}

In the same sense as before this method therefore \emph{allows us to reproduce the exact eigenvalues of the Hamiltonian}. In this sense it is exact. 

Of course, this first truncated expression is not exact for higher spins $S$ because couplings to the non-physical states reappear. We can obtain exact expressions also for $S>1/2$ by setting $n_{\mathrm{max}}=2S$. Similarly to the $S=1/2$ case these expressions reproduce all physical matrix elements. The proof is given in the Appendix~\ref{proof_no_coupling_to_non_phys}, but is essentially the same as for spin $1/2$. A list of explicit expressions for spin operators up to $S=3$ are given in Appendix~\ref{app:higherspin}.

\subsection{Commutator properties and exactness for the improved expansion}
One may ask what happens to commutators. Here the spin 1/2 case again is instructive,
\begin{equation}
	[S^+,S^-]\approx 2\hbar S^z-3 h^2 \left(S \left(2 \sqrt{4-\frac{2}{S}}-4\right)+1\right) a^{\dag^2}a^2.
\end{equation}

While the commutator is not exactly reproduced we can immediately recognize that this is not important because the extra term ${a^\dag}^2a^2$ does not couple unphysical and physical parts of the Hilbert space and solely affects the unphysical parts. It is therefore of no physical consequence. 

This additional term often was understood as rendering the expressions for spin operators approximate \cite{Lindgard1974}. After all, the most commonly used criterion for ruling out if an operator can be expressed in a certain way is by checking the commutation relations. Here we stress that this criterion can be softened. Namely it can be enough to reproduce commutation relations up to the addition of a term that acts solely in the unphysical part of the Hilbert space and does not couple to the physical part of the Hilbert space.

In some cases more stringent exactness criteria have been applied, such as requiring that all non-physical matrix elements vanish \cite{Zhou_1999}. This type of criteria can simplify formal quantum statistical treatments, since one does not have to be careful about excluding non-physical states in sums over states. However, in practice these approaches are cumbersome because the associated expansions are infinite and more complicated. Therefore this is only an advantage at the purely formal level.

\subsection{Additional properties of the expansion and comparison to other expansions}
One may wonder how this expansion compares to a more conventional Dyson Maleev expansion with  $S^+=\hbar a$ and $S^-=\hbar a^\dag(2S-a^\dag a)$. Our method has the advantage that $S^+$ and $S^-$ are treated on the same footing and therefore are related by conventional Hermitian conjugation. This guarantees that the approach will not break hermiticity in the conventional sense, unlike the Dyson-Maleev expansion.

Next one may wonder if an additional perturbative expansion around classical spin configurations may be stacked on top of the expansion as it is done for the more conventional $1/S$ expansion in non-linear spinwave theory \cite{RevModPhys.85.219}. One may therefore be tempted to identify $\delta=\left(\sqrt{1-\frac{1}{2 S}}-1\right)$ in Eq. \eqref{trunc_splus}  as an expansion parameter since it corresponds to fluctuation corrections around a classical ground state. That is, one would write $S^+\approx \hbar \sqrt{2S}\left[1+\delta a^\dag a\right]a$.  This, however, is not possible and becomes clear if one considers that $S^z=1/2[S^+,S^-]$. Then, one can write
\begin{equation}
\begin{aligned}
S^z=&\frac{1}{2}[S^+,S^-]\approx \underbrace{S+2S\left( 2 \delta +\delta ^2 \right) a^{\dagger } a}_{S^z}+3 \delta ^2 S \left(a^{\dagger }\right)^2 a^2.
\end{aligned}
\end{equation}
 We find that the physical part of $S^z$ has contributions from different orders of $\delta$. This of course means that any expansion in $\delta$ will treat $S^z$ and $S^{x,y}$ on unequal footing, even at low orders in such an expansion. This, for instance, will result in an unphysical breaking of symmetries in a Heisenberg model or similar even at the lowest order expansion. Therefore, $\delta$ cannot be used as an expansion parameter.  Additionally, there is no other obvious choice of expansion parameter and ad-hoc expansions in powers of $a$ also lead to unphysical results in non-linear spin-wave theories. Therefore, it seems that the expansion does not allow for an additional perturbative expansion in terms of fluctuations around a classical spin configuration. A mean field theory treatment must include all the terms that are needed to accurately describe spin $S$ for each operator $S^+$ and $S^-$.
 
 \subsection{Symmetries and exact properties in the improved expansion}
 
 To study symmetries in the new expansion we consider the Hamiltonian for the Heisenberg model with easy-plane single-ion anisotropy,
 \begin{equation}
 	H=\sum_i \left[ J\vect S_i \cdot \vect S_{i+1}+D(S^x_i)^2 \right].
 \end{equation}
 
 Let us first recognize that for $S=1/2$ the single-ion anisotropy $(S^x_i)^2$ should result in a trivial number $(S^x_i)^2=1/4$ that does not affect the spin-wave excitation spectrum. However, in the usual Taylor expansion with $S^+\approx a-\frac{1}{2}a^\dag a^2$ one finds that 
  \begin{equation}
 \begin{aligned}
 (S^x)^2&=\frac{1}{4}+\frac{1}{16} \left(2 {a^{\dagger }}^2+2 a^{\dagger } a-2 a^{\dagger } a^3-3 {a^{\dagger }}^2 a^2\right.\\
 &\left.+{a^{\dagger }}^2 a^4-2 {a^{\dagger }}^3 a+2 {a^{\dagger }}^3 a^3+{a^{\dagger }}^4 a^2+2 a^2\right),
 \end{aligned}
 \end{equation}
 which has unphysical contributions in the physical part of the Hilbert space, e.g. $a^\dagger a$. In other words, the Taylor expansion introduces unphysical artifacts. 
 
 In the new expansion for $S=1/2$, however, we have $S^+=a-a^\dag a^2$ and find that
 
 \begin{equation}
 	\begin{aligned}
 	 &(S^x)^2=\frac{1}{4}+\frac{1}{4} \left( {a^{\dagger }}^2 a^2+{a^{\dagger }}^2 a^4+2 {a^{\dagger }}^3 a^3+{a^{\dagger }}^4 a^2\right).
 	 \end{aligned}
 \end{equation}
  The additional non-constant terms we find have non-zero contributions only in the non-physical part of the Hilbert space, and do not couple to the physical part of the Hilbert space. This can easily be verified explicitly by computing the operator in the occupation number basis using Eq.~\eqref{eq:splusnew:matrix}. The non-physical terms are therefore of no consequence for physical states, and could just as well be dropped. This means that the new expansion properly reproduces the fact that $(S^x_i)^2$ contributes only a trivial scalar for spin $1/2$.
 
 Next we recall that the Hamiltonian  is symmetric with respect to the symmetry generated by the generator $g=\sum_i S_i^x,$ i.e. $C=[H,g]=0$. Again we will only check spin $1/2$ for simplicity, but similar results will hold for higher spins. Let us first see what happens if we use the usual Taylor expansion approach to compute the commutator. We find that 
 \begin{equation}
 	\begin{aligned}
 	C&=\sum_i\frac{1}{16} a_{i+1}^{\dagger } a_{i+1} \left(2 a_i^{\dagger }+a_i^{\dagger } a_i^2-{a_i^{\dagger }}^2 a_i-2 a_i\right)\\
 	&+\frac{3}{32} {a_{i+1}^{\dagger }}^2 a_{i+1}^2 \left(2 a_i^{\dagger }+a_i^{\dagger } a_i^2-{a_i^{\dagger }}^2 a_i-2 a_i\right)\\
 	&+(i)\leftrightarrow(i+1)
 	\end{aligned},
 \end{equation}
 where $(i)\leftrightarrow(i+1)$ is a shorthand for the same terms with $i$ and $i+1$ switched. Here we can see that the operators in the first line couple the physical two site states in $A_p=\{\left|0\right\rangle_i\left|0\right\rangle_{i+1},\left|1\right\rangle_i\left|0\right\rangle_{i+1},\left|0\right\rangle_i\left|1\right\rangle_{i+1},\left|1\right\rangle_i\left|1\right\rangle_{i+1}\}$ to non-physical two-site states in $A_{np}=\left\{\left|n\right\rangle_{i}\left|m\right\rangle_{i+1}| (n>1)\lor (m>1) \right\}$, which are the states where at least one of the two sites is more than single occupied. The symbol $\lor$ denotes the inclusive ``$\mathrm{or}$'' (disjunction) operator.

For the new expansion, on the other hand, we find that
\begin{equation}
	\begin{aligned}
	&C=\frac{3}{4}\sum_i\left({a_{i+1}^{\dagger }}^2 a_{i+1}^2 \left[a_i^{\dagger }-a_i+a_i^{\dagger } a_i^2-{a_i^{\dagger }}^2 a_i\right]\right.\\
	&\left.+{a_i^{\dagger }}^2 a_i^2 \left[a_{i+1}^{\dagger }-a_{i+1}+a_{i+1}^{\dagger } a_{i+1}^2-{a_{i+1}^{\dagger }}^2 a_{i+1}\right]\right).
	\end{aligned}
\end{equation}

From here one may observe the term ${a_{j}^{\dagger }}^2 a_{j}^2$ to see that only matrix elements for the unphysical states $A_{np}$ will be non-zero. The operator also does not couple to the physical states in $A_p$. We can hence conclude that, unlike the Taylor expansion, the bosonic expressions for the spin operators in the new expansion do not break symmetries present in the original spin operator language.

\section{Conclusion}

We were able to demonstrate the surprising result that the square-root of an operator $\sqrt{\hat O}$ may be expanded in an integer power series around $\hat O=0$. We believe that the approach can be usefully applied to other operator square-roots in theoretical physics and that the observation is useful for finding better expansions of other operator functions where a Taylor expansion fails.

The methods described in this paper allowed us to find a significant non-perturbative improvement on the Taylor expansion for the Holstein-Primakoff realization of spin operators. We expect these results to be useful to better treat spin models in different mean field approaches if there is no clear classical spin configuration around which one could expand. We therefore hope that the approach will prove useful for the study of spin liquid phases.

\acknowledgments
We thank C. D. Batista and G. Marmorini for useful discussions. M.V. and G.A.F. gratefully acknowledge partial support from the National Science Foundation through the Center for Dynamics and Control of Materials: an NSF MRSEC under Cooperative Agreement No. DMR-1720595, and also from NSF Grant No. DMR-1949701. PL was supported by the Scientific Discovery through Advanced Computing (SciDAC) program funded by the US Department of Energy, Office of Science, Advanced Scientific Computing Research and Basic Energy Sciences, Division of Materials Sciences and Engineering. SO was supported by U.S. DOE, Office of Science, Basic Energy Sciences, Materials Sciences and Engineering Division.

\bibliography{literature}
\appendix
\section{More generic case for operator square-root}
\label{general_expansion}
An operator square-root 
\begin{equation}
	S=\sqrt{O_1+\delta O_2}
\end{equation}
with small $\delta\ll1$ can generically be treated as follows.
One may write 
\begin{equation}
	S^2=O_1+\delta O_2,
\end{equation}
and make the ansatz $S=\sqrt{O_1}+\delta S_1$ to find
\begin{equation}
	S_0S_1+S_1S_0=O_2.
\end{equation}

This equation is a Lyapunov equation, which for $S_0$ Hermitian with positive spectrum can be solved for $S_1$ as
\begin{equation}
	S_1=\int_0^\infty dt e^{-S_0t}O_2 e^{-S_0t},
\end{equation}
which can be seen if it is inserted in the equation above and a chain rule for differentiation is used. Therefore to linear order one finds
\begin{equation}
	\sqrt{O_1+\delta O_2}\approx\sqrt{O_1}+\delta \int_0^\infty dt e^{-\sqrt{O_1}t}O_2 e^{-\sqrt{O_1}t}.
\end{equation}

It is easy to see that this, with the assumption $[O_1,O_2]=0$ we made earlier in the text, reduces to a Taylor series result. This result, however, is much more cumbersome and we will therefore not work on it further.

If we set $O_1=O(s)$, $O_2= \frac{dO_1}{ds}$ and $\delta=ds$ we find that a derivative of the square-root map is 
\begin{equation}
	\frac{d}{ds}\sqrt{O(s)}=\int_0^\infty dt e^{-\sqrt{O(s)}t}\frac{dO(s)}{ds} e^{-\sqrt{O(s)}t}.
\end{equation}

\section{Reproducing spin-\texorpdfstring{$S$}{S} operators exactly}
\label{proof_no_coupling_to_non_phys}
Let us first prove that truncating \eqref{new_expans_spin} at $n_\mathrm{max}=2S$ produces terms that don't couple to the non-physical parts of the Hilbert space. In the number basis we find that
\begin{equation}
\langle m|S^+|n\rangle=\hbar \sqrt{2S}\delta_{m,n-1}\sqrt{n}\sum_{l=0}^{2S}Q_l\frac{m!}{(m-l)!}
\end{equation}

The only \textit{a priori} non-zero matrix element that could couple physical and unphysical parts of the Hilbert space is $n=2S+1$ and $m=2S$. It is zero if 
\begin{equation}
\sum_{l=0}^{2S}Q_l\frac{2S!}{(2S-l)!}=0,
\end{equation}
which we checked explicitly for spins $S=1/2,...,16$ using Mathematica and expect to be true in general.

Furthermore we also checked explicitly that the other non-zero matrix elements for the physical couplings i.e. $n<2S+1$ and $m=n-1$ agree with the ones given by the exact Holstein-Primakoff expansion.
\begin{equation}
\langle m|S^+|n\rangle=\hbar \sqrt{2S}\delta_{m,n-1}\sqrt{1-\frac{m}{2S}}\sqrt{n}.
\end{equation}
That is we just needed to show that
\begin{equation}
\sum_{l=0}^{2S}Q_l\frac{m!}{(m-l)!}=\sqrt{1-\frac{m}{2S}}
\end{equation}
for all $m\in \mathbb{N}^+/2$ and $m<2S$.
Again using Mathematica we found this to hold at the minimum up to spin $S=16$ and expect it to be true generally.

\section{Explicit expressions for higher spin \texorpdfstring{$S$}{S}}
\label{app:higherspin}
In this appendix we give exact expressions for spin operators of spins up to $S=3$. For this one first has to solve \eqref{new_expans_spin} for the different $Q_n$ given below
\begin{widetext}
\begin{equation}
	\begin{aligned}
	&Q_0=1\\
	&Q_1=\sqrt{1-\frac{1}{2 S}}-1\\
	&Q_2=\frac{1}{2} \left(-\sqrt{4-\frac{2}{S}}+\sqrt{\frac{S-1}{S}}+1\right)\\
	&Q_3=\frac{1}{12} \left(3 \sqrt{4-\frac{2}{S}}+\sqrt{4-\frac{6}{S}}-6 \sqrt{\frac{S-1}{S}}-2\right)\\
	&Q_4=\frac{\sqrt{S-2}+6 \sqrt{S-1}+\sqrt{S}-2 \sqrt{4 S-6}-2 \sqrt{4 S-2}}{24 \sqrt{S}}\\
	&Q_5=\frac{-10 \sqrt{S-2}-20 \sqrt{S-1}-2 \sqrt{S}+\sqrt{4 S-10}+10 \sqrt{4 S-6}+5 \sqrt{4 S-2}}{240 \sqrt{S}}\\
	&Q_6=\frac{\sqrt{S-3}+15 \sqrt{S-2}+15 \sqrt{S-1}+\sqrt{S}-3 \sqrt{4 S-10}-10 \sqrt{4 S-6}-3 \sqrt{4 S-2}}{720 \sqrt{S}}
	\end{aligned}
\end{equation}
\end{widetext}

This result may now be inserted into $S_+= \hbar \sqrt{2S}\left[\sum_{n=0}^{2S} Q_n a^{\dag^n} a^n\right]a$ and setting the appropriate values for $S$ to find exact expressions for spin operators up to spin $S=3$ given below
\begin{widetext}
	\begin{equation}
	\hspace*{-0.5cm}
		\begin{aligned}
	&\underline{S=\frac{1}{2}}\text{: }\frac{S^+}{\hbar}=a-a^\dag a^2\\
	&\underline{S=1}\text{: }\frac{S^+}{\hbar}=\sqrt{2}a+\left(1-\sqrt{2}\right)  a^\dag a^2 + \left(\frac{1}{\sqrt{2}}-1\right) {a^{\dag}}^2a^3 \\
	&\underline{S=\frac{3}{2}}\text{: }\frac{S^+}{\hbar}=\sqrt{3}a+\left(\sqrt{2}-\sqrt{3}\right)  a^\dag a^2+\frac{1}{2} \left(1-2 \sqrt{2}+\sqrt{3}\right) {a^{\dag}}^2a^3+\frac{1}{6} \left(\sqrt{21-6 \sqrt{6}}-3\right) {a^{\dag}}^3a^4\\
	&\underline{S=2}\text{: }\frac{S^+}{\hbar}=2a+ \left(\sqrt{3}-2\right) a^\dag a^2+\left(1-\sqrt{3}+\frac{1}{\sqrt{2}}\right)  {a^{\dag}}^2a^3+\frac{1}{6} \left(3 \sqrt{3}-3 \sqrt{2}-1\right) {a^{\dag}}^3a^4\\
	&\hspace*{2cm}+\frac{1}{12} \left(3 \sqrt{2}-2 \sqrt{3}-1\right) {a^{\dag}}^4a^5\\
	&\underline{S=\frac{5}{2}}\text{: }\frac{S^+}{\hbar}=\sqrt{5}a+\left(2-\sqrt{5}\right) a^\dag a^2+\frac{1}{2} \left(\sqrt{3}+\sqrt{5}-4\right) {a^{\dag}}^2a^3+\frac{1}{6} \left(\sqrt{2}-3 \sqrt{3}-\sqrt{5}+6\right) {a^{\dag}}^3a^4\\
	&\hspace*{2.2cm}+\frac{1}{24} \left(\sqrt{5}-4 \sqrt{2}+6 \sqrt{3}-7\right) {a^{\dag}}^4a^5+\frac{1}{120} \left(10 \sqrt{2}-10 \sqrt{3}-\sqrt{5}+5\right) {a^{\dag}}^5a^6 \\
	&\underline{S=3}\text{: }\frac{S^+}{\hbar}=\sqrt{6}a+\left(\sqrt{5}-\sqrt{6}\right) a^\dag a^2+\left(1-\sqrt{5}+\sqrt{\frac{3}{2}}\right) {a^{\dag}}^2a^3+\frac{1}{6} \left(\sqrt{3}+3 \sqrt{5}-\sqrt{6}-6\right) {a^{\dag}}^3a^4\\
	&\hspace*{2cm}+\frac{1}{24} \left(\sqrt{2}-4 \sqrt{3}-4 \sqrt{5}+\sqrt{6}+12\right) {a^{\dag}}^4a^5+\frac{1}{120} \left(10 \sqrt{3}-5 \sqrt{2}+5 \sqrt{5}-\sqrt{6}-19\right) {a^{\dag}}^5a^6\\
	&\hspace*{2cm}+\frac{1}{720} \left(15 \sqrt{2}-20 \sqrt{3}-6 \sqrt{5}+\sqrt{6}+24\right) {a^{\dag}}^6a^7\\
		\end{aligned}
	\end{equation}
\end{widetext}

\end{document}